\documentclass[12pt]{article}
\usepackage{graphicx}
\usepackage{siunitx}
\usepackage{amsmath}
\usepackage{subcaption}

\newcommand\pubnumber{NuPhys2018-Vinning}
\newcommand\pubdate{\today}

\def\warwick{Department of Physics\\
University of Warwick, Coventry CV4 7AL, United Kingdom}
\def\support{\footnote{Work supported by the UK Science and Technology Facilities Council.}}
\def\collab{\footnote{On behalf of the Hyper-Kamiokande proto-collaboration.}}
\def\Title#1{\begin{center} {\Large #1 } \end{center}}
\def\Author#1{\begin{center}{ \sc #1} \end{center}}
\def\Address#1{\begin{center}{ \it #1} \end{center}}

\newcommand\pubblock{\rightline{\begin{tabular}{l} \pubnumber\\
         \pubdate  \end{tabular}}}
\newenvironment{Abstract}{\begin{quotation}  }{\end{quotation}}
\newenvironment{Presented}{\begin{quotation} \begin{center} 
             PRESENTED AT\end{center}\bigskip 
      \begin{center}\begin{large}}{\end{large}\end{center} \end{quotation}}
\def\Acknowledgements{\bigskip  \bigskip \begin{center} \begin{large}
             \bf ACKNOWLEDGEMENTS \end{large}\end{center}}

\begin{document}
\begin{titlepage}
\pubblock

\vfill
\Title{The Narrow-beam Diffuser Subsystem of a Prototype Optical Calibration System for the Hyper-Kamiokande Detector}
\vfill
\Author{ William G. S. Vinning\collab$^,$\support}
\Address{\warwick}
\vfill
\begin{Abstract}

The Hyper-Kamiokande neutrino detector is set to begin construction in 2020, succeeding Super-Kamiokande as the world's largest water Cerenkov detector. Research and development are well underway for an integrated light injection system for Hyper-Kamiokande which will provide in-situ monitoring of photo-sensor responses and water transparency. In summer 2018, optical hardware forming an iteration of this system was installed in Super-Kamiokande. We present details of the narrow-beam diffuser hardware and testing procedures, in addition to a brief summary of the installed light injection system.

\end{Abstract}
\vfill
\begin{Presented}
NuPhys2018, Prospects in Neutrino Physics

Cavendish Conference Centre, London, UK, December 19--21, 2018
\end{Presented}
\vfill
\end{titlepage}
\def\thefootnote{\fnsymbol{footnote}}
\setcounter{footnote}{0}

\section{Introduction}

Hyper-Kamiokande (HK)\cite{Abe:2018uyc} is set to become the world's largest underground water Cerenkov detector, succeeding Super-Kamiokande (SK) in target mass by a further order of magnitude. Given the historic success and proven technology of SK, the general detector design of HK will draw heavily from its predecessor. Within the domain of detector calibration particularly, the techniques laid out by SK\cite{Abe:2013gga} are sufficient and expandable to HK with little issue. However, considering the sheer scale of HK, ease of deployment and low-maintenance operation are desirable qualities of future calibration hardware. Hence, an integrated light injection system is being developed by UK groups to perform both inter-PMT (gain, timing and multi-photon) and optical (absorption and scattering) calibrations in-situ.

In coordination with refurbishment efforts in the summer of 2018, optical hardware constituent of this system was installed in the SK inner-detector. The experience gained from this deployment is anticipated to greatly instruct the design of a final system for HK. This document will outline the narrow-beam diffuser hardware comprising the optics of this system only. In Section 2, the light injection system installed is briefly described. The narrow-beam diffuser hardware is detailed in Section 3, accompanied by an insight into lab testing procedures.

\section{Overview of the Light Injection System}

\begin{figure}[h!]
    \centering
    \includegraphics[width=0.31\linewidth]{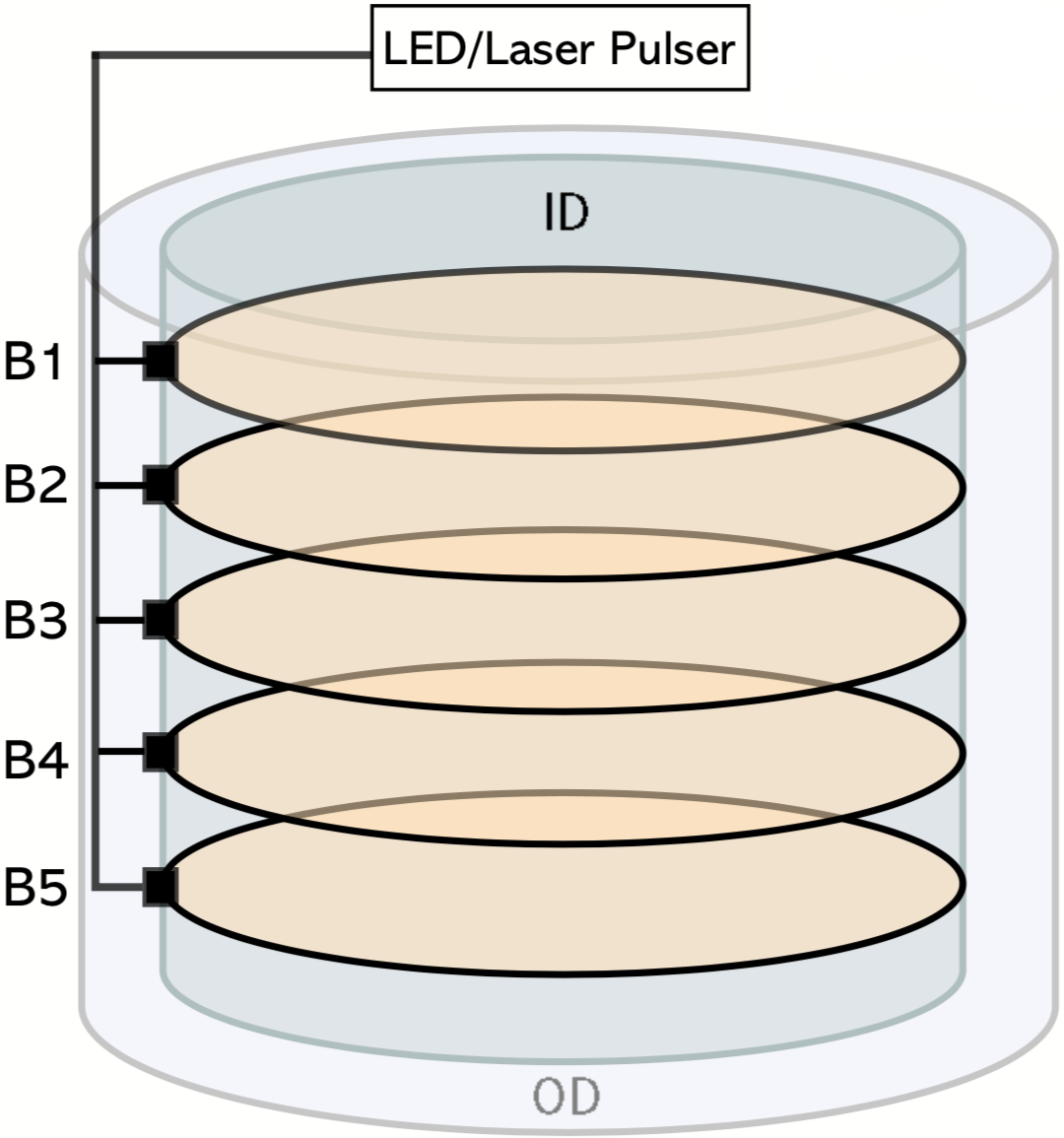}
    \caption{Light injection points (black squares) forming the calibration system installed in Super-Kamiokande (not to scale).}
    \label{fig:sk}
\end{figure}

Laser/LED light is transported into the tank via \SI{200}{\micro\metre} core graded-index optical fibres. Two kinds of optical diffuser establish each injection point, which each shape injected light to produce characteristic sources in the tank. The SK system is depicted in Figure \ref{fig:sk}, in which five injection points, facing towards the centre of the inner detector, are installed at increasing depth on the PMT structure. 

One of these diffusers achieves a wide-angle beam, illuminating large numbers ($\sim$100's) of photo-sensors uniformly to characterise relative PMT properties. The other attains a narrow-beam, projecting a beam with diameter no larger than five inner-detector PMT spacings for the use of monitoring water transparency. These two optical assemblies, in addition to a bare fibre for validation purposes, are fixed to a mounting bracket which in turn is secured to the inner-detector structure. The fibre connection from the optics to source was made in the tank via custom water-tight connection boxes designed to withstand SK water pressures long-term.

\section{The Narrow-beam Diffuser}

\subsection{Transparency monitoring strategy}

\begin{figure}[h!]
    \centering
    \includegraphics[width=0.45\linewidth]{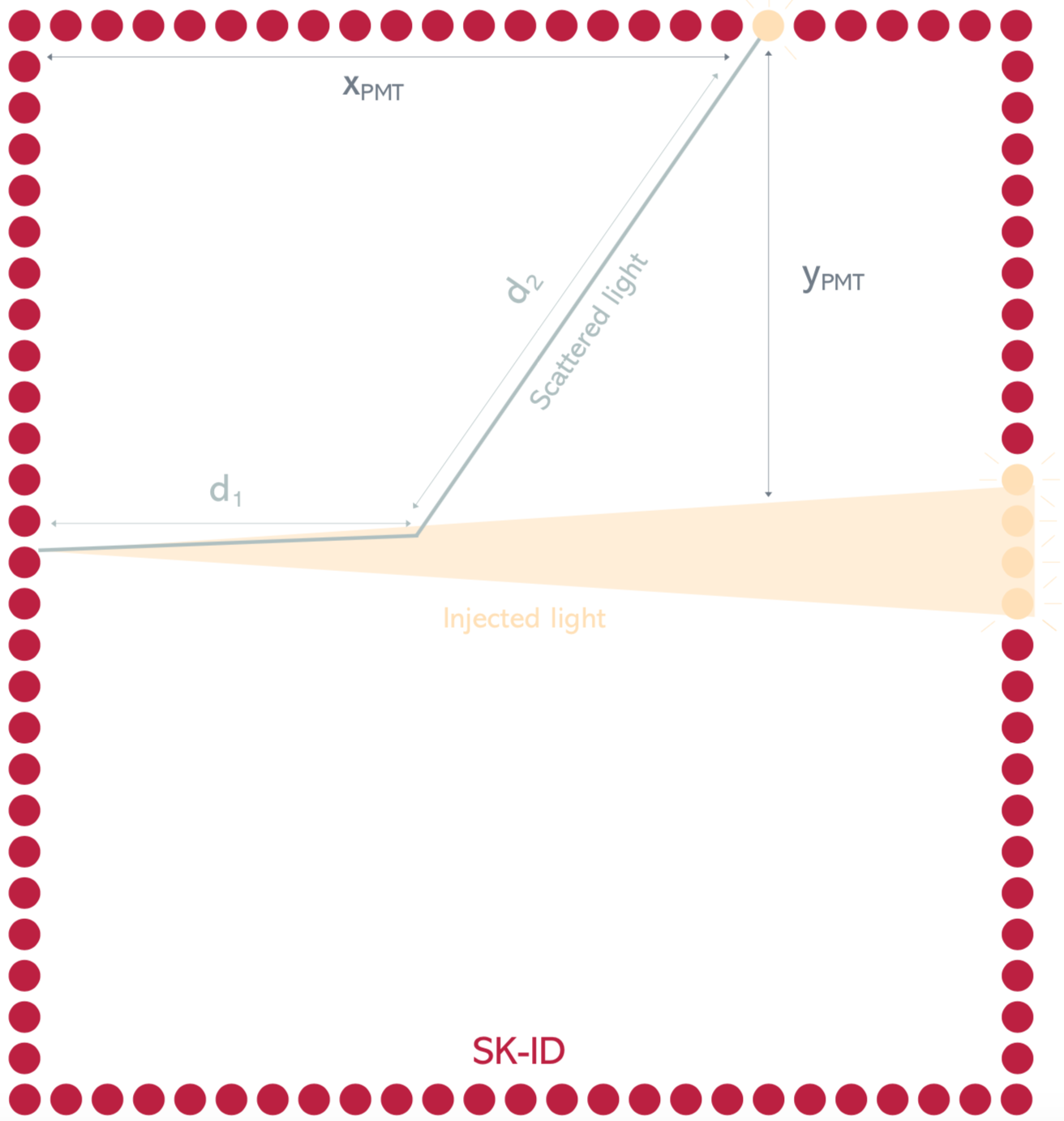}
    \caption{Photon scattering length derivation with a narrow-beam.}
    \label{fig:scattering}
\end{figure}

The narrow-beam diffuser produces a source purposed to directly measure scattering and absorption lengths, which are quantities useful for photon tracking calibrations. To achieve this, the interaction vertex of scattered photons may be reconstructed from hit PMT coordinates and timing information. As given in Equation \ref{eq:scatter} and illustrated in Figure \ref{fig:scattering}, this may be achieved by applying simple trigonometry.

\begin{equation}
  \begin{gathered}
    c\Delta t = n(d_1 + d_2) \\
    d^2_2 = (x_{PMT} - d_1)^2 + y^2_{PMT}
  \end{gathered}\label{eq:scatter}
\end{equation}

\subsection{Hardware}

\begin{figure}[h!]
    \centering
    \includegraphics[width=\linewidth]{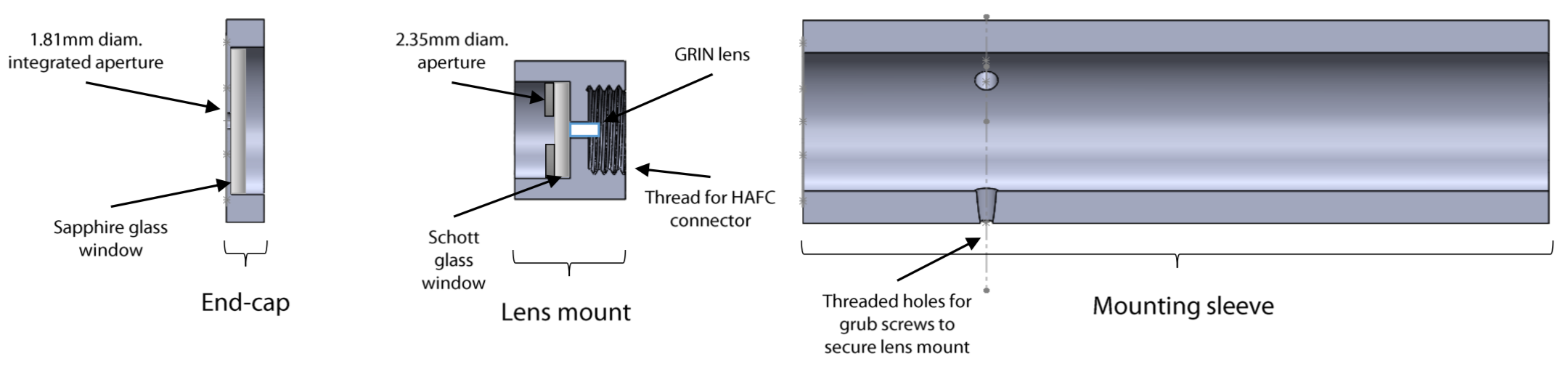}
    \caption{Schematic comprising the narrow-beam diffuser assembly.}
    \label{fig:schem}
\end{figure}

The collimation provided by the narrow-beam diffuser is generated by a 1/4 pitch gradient-index lens succeeded with a pair of apertures downstream. The optics are mounted inside a stainless steel assembly as depicted in Figure \ref{fig:schem}. 

To ensure beam stability, the optical fibre is mechanically secured to the body of the lens mount with a suited connector, the ferrule of which in turn holds the lens in place against a glass window. Rays inside a radial gradient-lens travel in a sinusoidal trajectory due to a quadratic dependence on refractive index with lens radius. For a 1/4 pitch lens, rays exit the lens at a phase of $\pi/2$ in the oscillation cycle, creating a perfectly collimated beam in a point source scenario. The opening angle of an output beam focused by a gradient index lens is dependent on the beam diameter of an aligned input source located immediately at the lens face.

The lens mount is fixed inside a cylindrical sleeve onto which an end-cap is fitted to seal the system. A \SI{1.81}{\milli\metre} diameter aperture integrated into the end-cap filters the beam of off-axis flux. The distance between the end-cap and was set individually for each unit such that the half-opening angle of the full-width half-maximum output cone was measured to be \SI[separate-uncertainty, multi-part-units = single]{1.8+-0.1}{\degree} in air. A homebrew photon transport Monte Carlo simulation informed the development process greatly.

\subsection{Testing procedures}

Beam measurements were made with a test-stand consisting of a monochrome CMOS camera mounted onto a system of high-precision linear stages, coupling a $\sim$\SI{450}{\nano\meter} LED to the diffuser. The transverse and longitudinal profiles of each diffuser were characterised prior to deployment for later use in analysis of calibration data. Each unit also was pressure-tested up to \SI{6}{\bar} with no failures reported. 

\section{Summary}

\begin{figure}[htp]
            \centering
            \begin{subfigure}{.48\textwidth}
              \centering
              \includegraphics[width=\linewidth]{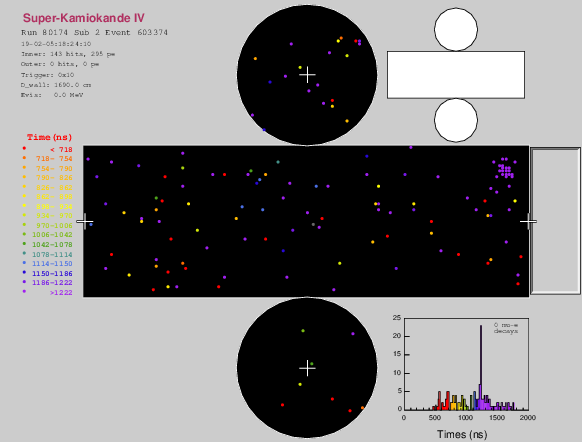}
              \caption{B1 (topmost) injector.}
              \label{fig:b1}
            \end{subfigure}%
            \hfill
            \begin{subfigure}{.48\textwidth}
              \centering
              \includegraphics[width=\linewidth]{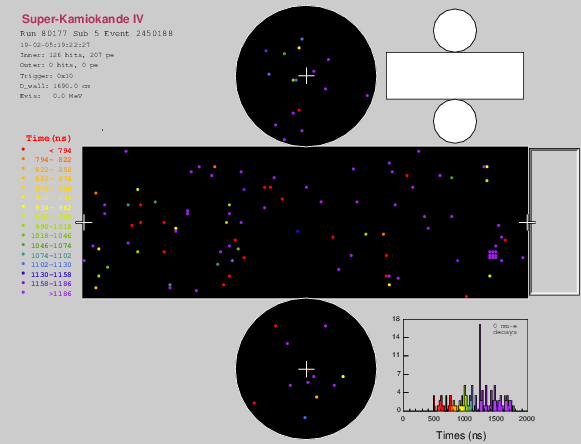}
              \caption{B4 (second from bottommost) injector.}
              \label{fig:b2}
            \end{subfigure}
            \caption{Timing event displays from the narrow-beam diffuser signal check.}
            \label{fig:eventdisplay}
        \end{figure}

A collection of light injectors as part of research and development efforts for HK is installed in SK. Installation of the electronics is ongoing at the time of writing. In February 2019, signals from all injectors were detected from coupling the optics to a \SI{405}{\nano\metre} laser source. Timing event displays for the narrow-beam diffuser from these test runs are given in Figure \ref{fig:eventdisplay}. The signal is evident from the peak in timing distributions and bunching of hits displayed on the barrel region.

The positive signal observed is encouraging, though a more technical insight into the system's performance will require further analysis and data-taking. The lessons learned from this experience will be greatly conducive to the next phase of development, in which a light injection system for HK will be finalised.

\vspace{-0.3cm}
\Acknowledgements

We thank all collaborators who contributed to installation efforts. We also extend our gratitude to Kamioka Observatory, the Super-Kamiokande collaboration and the Super-Kamiokande calibration group for their assistance with installation and data-taking efforts in 2018.

\end{document}